\documentclass[aps,onecolumn,groupedaddress,nofootinbib,amsmath,amssymb]{revtex4}
\usepackage{dsfont}
\usepackage{bm}
\begin{document}

\title{Local supersymmetric extensions of the
Poincar\'e and AdS invariant gravity}

\author{Mokhtar Hassa\"{\i}ne}\email{hassaine-at-inst-mat.utalca.cl}
\affiliation{Instituto de Matem\'atica y F\'{\i}sica, Universidad de
Talca, Casilla 747, Talca, Chile,}\affiliation{Laboratoire de
Math\'ematiques et de Physique Th\'eorique, Universit\'e de Tours,
Parc de Grandmont, Tours, France,}

\author{Mauricio Romo}\email{mromo-at-physics.ucsb.edu}
\affiliation{Department of Physics, University of California, Santa
Barbara, CA 93106, USA.}

\begin{abstract}

In all the odd dimensions which allow Majorana spinors, we consider
a gravitational Lagrangian possessing local Poincar\'e invariance
and given by the dimensional continuation of the Euler density in
one dimension less. We show that the local supersymmetric extension
of this Lagrangian requires the algebra to be the maximal extension
of the $\mathcal {N}=1$ super-Poincar\'e algebra. By maximal, we
mean that in the right hand side of the anticommutator of the
Majorana super charge appear all the possible ``central charges''.
The resulting action defines a Chern-Simons gauge theory for the
maximal extension of the super-Poincar\'e algebra. In these
dimensions, we address the same problem for the AdS invariant
gravity and we derive its supersymmetric extension for the minimal
super-AdS algebra. The connection between both models is realized at
the algebraic level through an expansion of their corresponding Lie
super algebras. Within a procedure consistent with the expansion of
the algebras, the local supersymmetric extension of the Poincar\'e
invariant gravity Lagrangian is derived from the super AdS one.
\end{abstract}
\maketitle

\section{Introduction}
Two of the main fundamental assumptions in general relativity are
the requirements of general covariance and the fact that the field
equations for the metric are of second order. In view of this, in
three and four dimensions it is natural to describe the spacetime
geometry by the Einstein-Hilbert action (with or without a
cosmological constant) while for dimensions greater than four, a
more general theory can be used. This fact has been first noticed by
Lanczos in five dimensions \cite {LAN} and later generalized by
Lovelock for any dimensions $d$ \cite{LOV} . The resulting theory is
described by the Lovelock Lagrangian, which is a $d-$form
constructed with the vielbein $e^a$, the spin connection
$\omega^{ab}$, and their exterior derivatives without using the
Hodge dual. The corresponding action contains the same degrees of
freedom than those in the Einstein-Hilbert action \cite{TEZ} and is
the most general low-energy effective theory of gravity derived from
string theory \cite{ZW}. The Lagrangian is a polynomial of degree
$[d/2]$ in the curvature two-form, $R^{ab} = d\,\omega^{ab} +
\omega^{a}_{\;c} \wedge \omega^{cb}$,
\begin{equation}
{\mathcal L}^{(d)} = \sum_{p=0}^{[d/2]}\alpha_p~ \epsilon_{a_1\cdots
a_d} R^{a_1a_2}\cdots R^{a_{2p-1}a_{2p}}e^{a_{2p+1}}\cdots  e^{a_d}
~, \label{LovLag}
\end{equation}
where the $\alpha_p$ are arbitrary dimensionful coupling constants
and where wedge products between forms are understood. The first two
terms in (\ref{LovLag}) are nothing but the cosmological constant
and the Einstein--Hilbert pieces. The action is invariant under the
local Lorentz rotations by construction. Interestingly enough, in
odd dimensions $d = 2n+1$, there is a particular choice of the
coefficients $\alpha_p$ that allows to extend the local Lorentz
symmetry into a local (anti) de Sitter or Poincar\'e symmetry. The
latter choice is simply achieved by choosing $\alpha_p =
\delta_{\,p}^n$. The resulting gravitational Lagrangian
$\mathcal{L}_P$ corresponds to the dimensional continuation of the
Euler density and is given by
\begin{eqnarray}
\mathcal{L}_P=\epsilon_{a_1\cdots a_{2n+1}}R^{a_1a_2}\cdots
R^{a_{2n-1}a_{2n}}e^{a_{2n+1}}. \label{Poincareinvgravity}
\end{eqnarray}
Note that $\mathcal{L}_P$ reduces to the Einstein-Hilbert action
only in three dimensions. The invariance of the gravitational
Lagrangian $\mathcal{L}_P$ under the local Poincar\'e translations
whose action on the dynamical fields is given by
\begin{eqnarray}
\delta e^{a}=D\lambda^a:=d\lambda^a+\omega^{a}_{\,b}\lambda^b,\qquad
\delta\omega^{ab}=0, \label{Poincaretransl}
\end{eqnarray}
is a direct consequence of the Bianchi identity. In what follows, we
refer to the gravitational Lagrangian (\ref{Poincareinvgravity}) as
the Poincar\'e invariant gravity Lagrangian.

Let us stress an important feature that will be our guiding
principle in the construction of the supersymmetric extension of the
Lagrangian $\mathcal{L}_P$. The gravitational Lagrangian
(\ref{Poincareinvgravity}) belongs to the class of Chern-Simons
gauge theories with Yang-Mills gauge symmetries. Indeed, the local
Poincar\'e translations (\ref{Poincaretransl}) correspond to a gauge
transformation
\begin{eqnarray}
\delta_{\lambda} A=d\lambda+[A,\lambda], \label{gt}
\end{eqnarray}
with parameter $\lambda=\lambda^aP_a$. Note that we have
parameterized the components of the gauge field $A$ in the adjoint
representation of the Poincar\'e group with generators $P_a$ and
$J_{ab}$ as
\begin{eqnarray}
A=\frac{1}{2}\omega^{ab}J_{ab}+e^aP_a. \label{Poincareconnection}
\end{eqnarray}
In addition, the Lagrangian $\mathcal{L}_P$ is a Chern-Simons form
for the Poincar\'e connection (\ref{Poincareconnection}) since its
exterior derivative satisfies
\begin{eqnarray}
d\mathcal{L}_P=\langle F^{n+1} \rangle.
\end{eqnarray}
Here $F$ is the field strength associated to the Poincar\'e
connection (\ref{Poincareconnection}) and the symbol $\langle \cdots
\rangle$ refers to a symmetric invariant $(n+1)-$tensor on the
Poincar\'e group whose only nonvanishing component is given by
\begin{eqnarray}
\langle J_{a_1a_2},\cdots
J_{a_{2n-1}a_{2n}},P_{a_{2n+1}}\rangle=\frac{2^n}{n+1}\epsilon_{a_1\cdots
a_{2n+1}}. \label{eps}
\end{eqnarray}

In three dimensions the Lagrangian (\ref{Poincareinvgravity})
corresponds to the Einstein-Hilbert Lagrangian and the questions
relative to its supersymmetric extension have already been answered
in the past. In this case, the supersymmetry is easily achieved by
introducing only a spinor field and the resulting action can be
viewed as a gauge theory for the super-Poincar\'e algebra
\cite{MSSK}. Later on, the local supersymmetric extension of the
gravitational Lagrangian (\ref{Poincareinvgravity}) has been
constructed in arbitrary odd dimensions \cite{BTZ}. The resulting
supersymmetric theories possess a rich geometrical structure encoded
by a fibre bundle structure and by the fact that the supersymmetry
closes off-shell without requiring the introduction of auxiliary
fields. In a generic odd dimension, the supersymmetric extension of
the Poincar\'e invariant gravity Lagrangian defines a Chern-Simons
gauge theory with gauge group identified with the super five brane
algebra \cite{BTZ}. This algebra is a nontrivial extension of the
Poincar\'e algebra which is spanned by the Poincar\'e generators
together with a $5-$form ``central charges'' $Z_{a_1\cdots a_5}$ and
complex fermionic generators $Q^{\alpha}$ and $\bar{Q}_{\alpha}$
whose anticommutator reads
\begin{eqnarray}
\{Q^{\alpha},\bar{Q}_{\beta}\}=-i(\Gamma^a)^{\alpha}_{\,\beta}P_a-
\frac{i}{5!}\left(\Gamma^{a_1\cdots
a_5}\right)^{\alpha}_{\,\beta}Z_{a_1\cdots a_5}. \label{5brane}
\end{eqnarray}
Note that the bosonic generator $Z_{a_1\cdots a_5}$ required by
supersymmetry is truly a central charge only in five dimensions
while in odd dimensions $d>5$, it does not commute with the Lorentz
rotation generator because of its Lorentz indices. For some reviews
on Chern-Simons supergravity, see e.g. \cite{JZ}.

In view of the work of Ref.~\cite{BTZ}, the supersymmetric extension
of the Poincar\'e invariant gravity (\ref{Poincareinvgravity}) seems
to confer a particular status on the super five brane algebra
(\ref{5brane}). However, in nine dimensions it has been shown that
the super-Poincar\'e algebra with a $U(1)$ central extension can
accommodate such construction \cite{HOT} while in eleven dimensions
a supersymmetric extension of the Poincar\'e invariant gravity
(\ref{Poincareinvgravity}) with Majorana spinors has been achieved
for the M-algebra \cite{HTZ},
\begin{eqnarray}
\Huge{\left\{Q, {Q}\right\}=(C\Gamma^a)
P_a+(C\Gamma^{ab})Z_{ab}^{(2)}+(C\Gamma^{a_1\cdots a_5})Z_{a_1\cdots
a_5}^{(5)}}. \label{M-algebra}
\end{eqnarray}
Here $Z_{ab}^{(2)}$ and $Z_{a_1\cdots a_5}^{(5)}$ are ``central
charges'' corresponding to the two types of extended objects that
couple to the Abelian $3-$form of standard eleven supergravity
\cite{CJS}. From an algebraic point of view, the M-algebra
(\ref{M-algebra}) corresponds to the maximal extension of the ${\cal
N}=1$ super-Poincar\'e algebra in the sense that in the right hand
side of the anticommutator of the Majorana super charge appear all
the possible ``central charges'' allowing by symmetry
\cite{vanProeyen}. These two results suggest that the super five
brane algebra does not fulfill all the possibilities. Another
interesting observation concerns the dimensions $3$ and $11$ which
precisely allow Majorana spinors. In fact, in three dimensions
(resp. in eleven dimensions), the supersymmetric extension of the
Poincar\'e invariant gravity Lagrangian can be constructed and the
resulting action is a gauge theory for the maximal extension of the
$\mathcal {N}=1$ super-Poincar\'e algebra \cite{MSSK} (resp. in
\cite{HTZ}). The purpose of this article is to extend these two
results in all the odd dimensions allowing Majorana spinors
$d=3\;(\mbox{mod}\,8)$. Indeed, we shall prove that the maximal
extension of the ${\mathcal N}=1$ super-Poincar\'e algebra is always
compatible with the construction of the local supersymmetric
extension of the Poincar\'e invariant gravity Lagrangian. In other
words, this means that it is always possible to construct a local
supersymmetric extension of the Lagrangian
(\ref{Poincareinvgravity}) such that the resulting action can be
viewed as a gauge theory for the maximal extension of the ${\mathcal
N}=1$ super-Poincar\'e algebra. As shown below, the supersymmetric
action can be written in a simple form as a trace and its invariance
under supersymmetry is a direct consequence of a Fierz
rearrangement.

In these particular dimensions, we also construct the local
supersymmetric extensions of the AdS gravity for which the gauge
group is identified with the minimal supersymmetric extension of the
anti-de Sitter (AdS) group. By minimal we mean the smallest super
algebra that contains the AdS generators.

It is well-known that the Poincar\'e algebra can be viewed as a
Wigner-In\"on\"u contraction of the (A)dS algebra. Since we are
dealing with supersymmetric extensions of these algebras, it is
natural to ask whether the maximal extension of the ${\mathcal N}=1$
super-Poincar\'e algebra can be obtained from the smallest super
algebra containing the AdS generators. In the current literature,
there exist various ways to obtain a Lie algebra from a given one,
as for example the Wigner-In\"on\"u contraction. In general, these
procedures are restrictive in the sense that the starting and
resulting algebras have necessarily the same dimension. In
\cite{deAzcarraga:2002xi}, de Azc\'arraga {\it et al.} have proposed
a consistent way of generating a Lie algebra whose dimension is
greater than the original one. This method originally considered by
Hatsuda and Sakaguchi \cite{HS} in a less general context consists
in expanding the Maurer-Cartan one-forms in powers of a real
parameter in such way that the Maurer-Cartan equations are satisfied
order by order, leading to a closed algebra at each order. Within
this process called {\it expansion}, the M-algebra has been derived
from the minimal extension of the eleven-dimensional AdS algebra
$\frak{osp(32\vert 1)}$ that has $55$ generators less
\cite{deAzcarraga:2002xi}. In our case, we will show that the
maximal extension of the ${\mathcal N}=1$ super-Poincar\'e algebra
can be obtained from the minimal extension of the AdS algebra
through the expansion method. In addition, we also address the
problem of the correspondence at the level of the supersymmetric
actions that means finding a process compatible with the expansion
of the Lie super algebras that permit to obtain the super Poincar\'e
theory from the super AdS one. In this connection, Segal in
\cite{S}, stated that if two physical theories are linked through a
limiting process then there should also exist a corresponding limit
between their underlying symmetry groups. In this optics and as a
first step, we consider the standard Wigner-In\"on\"u contraction
which is the natural option to obtain an interesting theory at the
vanishing cosmological constant limit. In this case, we show that
although the Wigner-In\"on\"u contraction gives rise to a consistent
theory at the vanishing cosmological constant limit, the resulting
action being decoupled from the vielbein and the gravitino is of
little interest. In contrast, the next order in the contraction
contains the gravitational Lagrangian (\ref{Poincareinvgravity}) but
does not define a supersymmetric Lagrangian. As we shall show below,
the terms in the next order can be made supersymmetric by exploiting
the possibility of adding to the spin connection a tensor under the
Lorentz group. This splitting of the spin connection, which turns
out to be equivalent to expanding the minimal extension of the AdS
algebra, has important consequences. Indeed, within this process the
gauge structure relative to the maximal extension of the ${\mathcal
N}=1$ super-Poincar\'e algebra naturally emerges from the gauge
structure of the minimal extension of the AdS algebra. This means
that this procedure brings all the dynamical fields required to form
a connection for the maximal extension of the ${\mathcal N}=1$
super-Poincar\'e algebra as well as their prescribed supersymmetric
transformations that can be viewed as gauge transformations.
Finally, the local supersymmetric extension of the Poincar\'e
invariant gravity for the maximal extension of the ${\mathcal N}=1$
super-Poincar\'e algebra is deduced from the supersymmetric AdS
Lagrangian.

The paper is organized as follows. In the next section, we construct
the local supersymmetric extension of the Poincar\'e invariant
gravity in odd dimensions allowing Majorana spinors. We show that
the resulting action can be viewed as a gauge theory with a gauge
group identified with the maximal extension of the ${\mathcal N}=1$
super-Poincar\'e algebra. In the third section, the same problem is
addressed for the AdS invariant gravity and the link between both
models is established through the expansion method. The
eleven-dimensional case, because of its interest in the context of
the M-theory will be reported elsewhere \cite{HTZ2}. Finally, in the
last section we summarize our results, give some comments and
present some open questions.

\section{Local supersymmetric extension of the Poincar\'e invariant gravity}
We restrict ourselves to the odd dimensions allowing Majorana
spinors $d=3+8k$ with $k\in\mathds{N}$. We rewrite the gravitational
Lagrangian ${\mathcal L}_P$ defined in (\ref{Poincareinvgravity}) by
using a trace expression over the $\Gamma-$matrices as
\begin{eqnarray}
{\mathcal L}_P=\mbox{Tr}\left[R\!\!\!\!\slash^{4k+1}
e\!\!\!\slash\right] \label{traceL}
\end{eqnarray}
where we have defined
$$
e\!\!\!\slash=e_a\Gamma^a,\qquad\qquad
R\!\!\!\!\slash=\frac{1}{2}R_{ab}\Gamma^{ab}.
$$
As seen in the introduction, the Lagrangian ${\mathcal L}_P$
possesses local Poincar\'e invariance and the latter, among other
interesting features, can be viewed as a gauge theory for the
Poincar\'e algebra. A natural way to construct a local
supersymmetric extension of ${\mathcal L}_P$ sharing the same
features is to impose that the extra fields required by the
supersymmetry also belong to a connection for some supersymmetric
extension of the Poincar\'e algebra. In doing so, we shall see that
the maximal extension of the ${\mathcal N}=1$ super-Poincar\'e
algebra naturally emerges in order to accommodate the extra fields
required by the supersymmetry and also to prescribe their correct
supersymmetric transformations. As the simplest tentative, we see
whether the ${\mathcal N}=1$ super-Poincar\'e algebra without
central charges can accommodate this construction. In this case, the
field content is just supplemented by the introduction of the
gravitino $\psi$ and the supersymmetric transformations obtained as
a gauge transformation (\ref{gt}) with parameter
$\lambda=\epsilon^{\alpha}Q_{\alpha}$ where $\epsilon$ is a
zero-form Majorana spinor read
$$
\delta e^{a}=\left(\bar{\epsilon}\Gamma^a\psi\right), \qquad \delta
\omega^{ab}=0,\qquad
\delta\psi=D\epsilon:=(d+\frac{1}{4}\omega_{ab}\Gamma^{ab})\epsilon.
$$
In this case, the variation of ${\mathcal L}_P$ (\ref{traceL}) under
these supersymmetric transformations can be canceled by a kinetic
term of the gravitino $\psi$ given by
$$
{\mathcal L}_{\psi}=-2^{k+1}\mbox{Tr}\left[R\!\!\!\!\slash^{4k}\;
(D\psi)\bar{\psi}\right].
$$
In details, we have
$$
\delta{\mathcal L}_{P}+\delta{\mathcal
L}_{\psi}=\mbox{Tr}\Big[R\!\!\!\!\slash^{4k+1}\Big((\bar{\epsilon}\Gamma_a\psi)\Gamma^a
-2^k(\epsilon\bar{\psi}-\psi\bar{\epsilon})\Big)\Big],
$$
and then using the following Fierz rearrangement in $d=3+8k$
\begin{eqnarray}
\epsilon\bar{\psi}-\psi\bar{\epsilon}=\frac{1}{2^k}(\bar{\epsilon}\Gamma_a\psi)\Gamma^a+\sum_{p\in
{\mathcal P}}
\frac{(-1)^{p+1}}{2^{k}p!}\left(\bar{\epsilon}\Gamma_{a_1\cdots
a_p}\psi\right)\Gamma^{a_1\cdots a_p} \label{Fr}
\end{eqnarray}
where the sum is over the set ${\mathcal P}$ defined by
\begin{eqnarray}
{\mathcal P}=\left\{p=2, 5\;(\mbox{mod}\,4)\quad \mbox{with}\quad
p\leq 4k+1\right\}.
 \label{setP}
\end{eqnarray}
Finally, we obtain
\begin{eqnarray}
\delta{\mathcal L}_{P}+\delta{\mathcal L}_{\psi}=-\sum_{p\in
{\mathcal P}}
\frac{(-1)^{p+1}}{p!}\mbox{Tr}\Big[R\!\!\!\!\slash^{4k+1}\Big(\left(\bar{\epsilon}\Gamma_{a_1\cdots
a_p}\psi\right)\Gamma^{a_1\cdots a_p} \Big)\Big]. \label{etape1}
\end{eqnarray}
Hence, we conclude that the standard super-Poincar\'e algebra is not
rich enough to ensure the off-shell supersymmetry of the action.
Nevertheless, it is simple to see that the variation (\ref{etape1})
can be canceled by introducing bosonic one-form fields that are
tensors of rank $p$, $b^{a_1\cdots a_p}_{(p)}$ with $p\in {\mathcal
P}$, and transform as $\bar{\epsilon}\Gamma_{a_1\cdots a_p}\psi$
under supersymmetry. Assuming that these extra fields belong to a
single connection, the only option is to consider an extension of
the Poincar\'e algebra spanned by the following set of generators
\begin{eqnarray}
\left\{J_{ab}, P_a, Q_{\alpha}, (Z_{a_1\cdots a_p})_{p\in{\mathcal
P}}\right\}
 \label{algebra}
\end{eqnarray}
where $Q_{\alpha}$ is the Majorana generator and the generators
$(Z_{a_1\cdots a_p})_{p\in{\mathcal P}}$ are Lorentz tensors of rank
$p$. In this case, the corresponding super connection is given by
\begin{eqnarray}
A=\frac{1}{2}\omega^{ab}J_{ab}+e^aP_a+\psi^{\alpha}Q_{\alpha}+\sum_{{p\in{\mathcal
P}}}\frac{1}{p!}b^{a_1\cdots a_p}_{(p)}Z_{a_1\cdots a_p}.
\label{connection}
\end{eqnarray}
In addition, in order to prescribe the correct gauge supersymmetric
transformations of the extra bosonic fields, the anticommutator of
the Majorana generator must be given by
\begin{eqnarray}
\left\{Q,Q\right\}=(C\Gamma^{a})P_{a}+\sum_{p\in{\mathcal P}}
\frac{1}{p!}(C\Gamma^{a_1\cdots a_p})Z_{a_1\cdots a_p}, \label{maj}
\end{eqnarray}
where the sum is over the set ${\mathcal P}$ defined previously
(\ref{setP}) and where $C$ is the antisymmetric charge conjugation
matrix. The algebra (\ref{maj}) is known as the maximal extension of
the ${\mathcal N}=1$ super-Poincar\'e algebra. This algebra is said
maximal since the left hand side is a $2^{[d/2]}\times 2^{[d/2]}$
real symmetric matrix, so the maximal number of algebraically
distinct charges that can appear on the right hand side is
$2^{[d/2]}\times (2^{[d/2]}+1)/2$, which is precisely the number of
components of $P_a$ and the different p-form ``central charges''
that appear in the right hand side. In eleven dimensions, this
algebra is commonly known as the M-algebra since it encodes many
important features of the M-theory.

The supersymmetry transformations of all the dynamical fields can be
read off as a gauge transformation of the connection
(\ref{connection}) for the algebra (\ref{maj}), and they are given
by
\begin{eqnarray}
& & \delta e^{a}=\left(\bar{\epsilon}\Gamma^a\psi\right),
\quad \delta\psi=D\epsilon\nonumber\\
& & \delta \omega^{ab}=0,\quad \delta b^{a_1\cdots
a_{p}}_{(p)}=\left(\bar{\epsilon}\Gamma^{a_1\cdots
a_{p}}\psi\right). \label{mpea}
\end{eqnarray}

Finally, the local supersymmetric extension of the Poincar\'e
invariant gravity Lagrangian in the odd dimensions $d=3+8k$ is found
to be
\begin{eqnarray}
{\mathcal
L}_P^{\mbox{\tiny{susy}}}=\mbox{Tr}\left[R\!\!\!\!\slash^{4k}\left(R\!\!\!\!\slash
\;\Big(e\!\!\!\slash+\sum_{p\in {\mathcal P}}
(-1)^{p+1}b_{(p)}\!\!\!\!\!\!\!\!\!\slash\;\;\;\Big)
-\left(D\psi\right)\bar{\psi}\right)\right],\label{pok}
\end{eqnarray}
where we have defined
$$
e\!\!\!\slash=e_a\Gamma^a,\quad
R\!\!\!\!\slash=\frac{1}{2}R_{ab}\Gamma^{ab},\quad
b_{(p)}\!\!\!\!\!\!\!\!\!\slash\quad=\frac{1}{p!}b^{a_1\cdots
a_p}_{(p)}\Gamma_{a_1\cdots a_p}.
$$
The invariance of (\ref{pok}) with respect to the supersymmetric
transformations (\ref{mpea}) can easily be checked with the use of
the Fierz rearrangement (\ref{Fr}).

Hence, in odd dimensions allowing Majorana spinors, a supersymmetric
extension of the Poincar\'e invariant gravity can be constructed for
which the resulting action is a gauge theory for the maximal
extension of the ${\mathcal N}=1$ super-Poincar\'e algebra.

\section{Local supersymmetric extensions of the AdS invariant gravity}

In this section, we shall be concerned with the supersymmetric
extension of the AdS invariant gravity in odd dimensions allowing
Majorana spinors, $d=3+8k$. We shall also establish a connection
between this theory and the Poincar\'e one derived previously at
some vanishing cosmological constant limit.

In odd dimensions, the Lorentz symmetry of the Lovelock theory
(\ref{LovLag}) can also be extended to a local AdS symmetry and the
resulting Lagrangian is given by
\begin{eqnarray}
{\mathcal L}_{\tiny{\mbox{AdS}}}^{(2n+1)}= \sum_{q=0}^{n}\frac{{n
\choose q}}{(2n+1-2q)}\,{\mathcal L}^{(q)}, \label{adsact}
\end{eqnarray}
where the Lagrangians ${\mathcal L}^{(q)}$ are defined by
\begin{eqnarray}
{\mathcal L}^{(q)}=\epsilon_{a_1\cdots a_d}R^{a_1a_2}\cdots
R^{a_{2q-1}a_{2q}}e^{a_{2q+1}}\cdots e^{a_d}. \label{lp}
\end{eqnarray}
The action (\ref{adsact}) defines a $(2n+1)-$Chern-Simons form of
the AdS group. The supersymmetric extensions of the AdS gravity
action (\ref{adsact}) have been constructed in three \cite{AT,GTW},
five \cite{CHA} and higher odd dimensions \cite{TZ}.

In the present case, we are concerned with the odd dimensions
allowing Majorana spinors, $d=3+8k$. For these particular
dimensions, van Holten and Von Proeyen  have derived the minimal
super algebras that contain the AdS generators by adding one
Majorana supersymmetry generator and by demanding the closure of the
full super algebra \cite{vanHolten:1982mx}. In particular, the
consequences of the $[P,Q,Q]$ Jacobi identity imply that the
anticommutator of the Majorana generator is given by
\begin{eqnarray}
\left\{Q,Q \right\}=(C\Gamma^a)P_a-\frac{1}{2}(C\Gamma^{ab})J_{ab}
+\sum_{p^{\prime}\in {\mathcal
P}^{\prime}}\frac{1}{p^{\prime}!}\left(C\Gamma^{a_1\cdots
a_{p^{\prime}}}\right)Z_{a_1\cdots a_{p^{\prime}}}, \label{superAdS}
\end{eqnarray}
where the sum is over the set ${\mathcal P}^{\prime}$ defined by
\begin{eqnarray}
{\mathcal P}^{\prime}=\left\{p^{\prime}=5, 6\;(\mbox{mod}\,4)\quad
\mbox{with}\quad p^{\prime}\leq 4k+1\right\}. \label{setPp}
\end{eqnarray}
We insist on the notation $p^{\prime}$ to stress the difference with
the extended super-Poincar\'e algebras (\ref{maj}) where the
membrane value $p=2$ is included. The super algebras described by
the anticommutation relation (\ref{superAdS}) is known as the
orthosymplectic group $\mbox{Osp}(2^{[d/2]}\vert 1)$. There exist
important differences between the minimal super AdS algebras
(\ref{superAdS}) and the maximal extensions of the super-Poincar\'e
algebras (\ref{maj}). For a fixed dimension $d$, the super algebra
$\mbox{Osp}(2^{[d/2]}\vert 1)$ has $d(d-1)/2$ less generators than
the algebra (\ref{maj}) owing to the fact that the Poincar\'e
``central charge'' $Z_{ab}$ is not a generator of the super AdS
algebra. Another important difference is the presence of the Lorentz
generator $J_{ab}$ in the right hand side of the anticommutation
relation (\ref{superAdS}) which in turns implies that the
supersymmetric AdS transformation of the spin connection does not
vanish (\ref{mpae}).

The supersymmetric extension of the AdS gravity (\ref{adsact}) in
dimension $d=3+8k$ can be constructed as follows. We define a
connection one-form $A$ in the adjoint representation of
$\mbox{Osp}(2^{[d/2]}\vert 1)$. Then we construct the Chern-Simons
form associated as follows:
\begin{eqnarray}
d{\mathcal L}_{\tiny{\mbox{AdS}}}^{\tiny{\mbox{susy}}}=\mbox{STr}
\left(F^{4k+2}\right), \label{AdSsusyaction}
\end{eqnarray}
where ``STr''  stands for the super trace and $F$ is the curvature
associated to the connection $A$. The supersymmetric transformations
read off as gauge transformations are given by
\begin{eqnarray}
& & {\delta} e^{a}=\left(\bar{\epsilon}\Gamma^a\psi\right),
\qquad {\delta}\psi=\nabla\epsilon\nonumber\\
& & {\delta}
\omega^{ab}=-\left(\bar{\epsilon}\Gamma^{ab}\psi\right),\qquad
{\delta} b^{a_1\cdots
a_{p^{\prime}}}_{(p^{\prime})}=\left(\bar{\epsilon}\Gamma^{a_1\cdots
a_{p^{\prime}}}\psi\right) \label{mpae}
\end{eqnarray}
where the covariant derivative now reads
$$
\nabla\epsilon=D\epsilon+(e_a\Gamma^a+\sum_{p^{\prime}}b_{(p^{\prime})}^{a_1\cdots
a_{p^{\prime}}}\Gamma_{a_1\cdots a_{p^{\prime}}})\epsilon.
$$

As we are dealing with a theory in presence of a negative
cosmological constant $\Lambda$, a natural question to ask is
whether the limiting case $\Lambda\to 0$ yields to interesting
features. Moreover, it is well-known that the Poincar\'e algebra can
be viewed as a Wigner-In\"on\"u contraction of the (A)dS algebra,
and so it is legitimate to ask whether the Poincar\'e supersymmetric
theories described in the previous section can be derived from the
super AdS ones at the zero cosmological constant limit. In eleven
dimensions, this problem has been considered in \cite{HTZ2} where a
generalization of the Wigner-In\"on\"u contraction has permitted to
link the AdS supersymmetric theory invariant under
$\mbox{Osp}(32\vert 1)$ with a gauge theory for the M-algebra. The
aim of this section is to show that the conclusions obtained in
eleven dimensions are still valid for all the odd dimensions
admitting Majorana spinors. More precisely, we first show that the
maximal Poincar\'e algebras (\ref{maj}) can be obtained from the
minimal super AdS algebras (\ref{superAdS}) through the expansion
method that permits to generate Lie algebra of higher dimensions
\cite{deAzcarraga:2002xi}. In addition, we shall see that the
implementation of this expansion on the dynamical fields also
permits to derive the supersymmetric extension of the Poincar\'e
invariant gravity (\ref{pok}) from the minimal super AdS theory
(\ref{AdSsusyaction}).

In order to point out the necessity of considering the expansion
method rather than a standard Wigner-In\"on\"u contraction, we first
operate a standard Wigner-In\"on\"u contraction on the
supersymmetric AdS action (\ref{AdSsusyaction}). The implementation
of this contraction on the AdS dynamical fields is given as usual by
\begin{eqnarray}
e^a\to \frac{1}{l}e^a,\qquad \omega^{ab}\to \omega^{ab},\qquad
b^{a_1\cdots a_{p^{\prime}}}_{(p^{\prime})}\to
\frac{1}{l}b^{a_1\cdots a_{p^{\prime}}}_{(p^{\prime})},\qquad
\psi\to \frac{1}{\sqrt{l}}\psi,\label{resmathcale}
\end{eqnarray}
where $l$ is a scaling parameter for the radius of the universe and
the zero cosmological constant limit corresponds to taking $l\to
\infty$ and where $p^{\prime}$ run over the set ${\mathcal
P}^{\prime}$ defined in (\ref{setPp}). On the other hand, since the
gauge parameter $\epsilon$ of the supersymmetric transformations
must also be rescaled as $\epsilon \to 1/\sqrt{l}\,\epsilon$, the
supersymmetric transformations $\bar{\delta}$ defined by
(\ref{mpae}) reduce to those associated to the extended
super-Poincar\'e algebra (\ref{mpea}) with the exception that the
bosonic field $b^{ab}_{(2)}$ is not present in the AdS theory
(\ref{setPp}). Operating the rescaling (\ref{resmathcale}) at the
level of the supersymmetric action (\ref{AdSsusyaction}), the latter
can be expanded as follows
\begin{eqnarray}
{\mathcal L}_{\tiny{\mbox{AdS}}}^{\tiny{\mbox{susy}}}&=&{\mathcal
L}^{\star}(\omega)+\frac{1}{l} \mbox{Tr}\Big[R\!\!\!\!\slash^{4k+1}
\Big(e\!\!\!\slash+\sum_{p^{\prime}\in {\mathcal P}^{\prime}}
(-1)^{p^{\prime}+1}b_{(p^{\prime})}\!\!\!\!\!\!\!\!\!\!\!\slash\;\;\;\Big)
-R\!\!\!\!\slash^{4k}
(D\psi)\bar{\psi}\Big)\Big]+o(l^{-2}),\nonumber\\
&=& {\mathcal L}^{(0)}+\frac{1}{l}{\mathcal L}^{(1)}+o(l^{-2})
\label{pokk}
\end{eqnarray}
where ${\mathcal L}^{(0)}={\mathcal L}^{\star}(\omega)$ is the
Lorentz Chern-Simons form which depends only on the spin connection
and which satisfies
\begin{eqnarray}
d{\mathcal
L}^{\star}(\omega)=\mbox{Tr}\left(R\!\!\!\!\slash^{4k+2}\right).
\label{LCS}
\end{eqnarray}
It is clear that in the limit $l\to\infty$, the supersymmetric
Lagrangian ${\mathcal L}_{\tiny{\mbox{AdS}}}^{\tiny{\mbox{susy}}}$
reduces to the Lorentz Chern-Simons form which is trivially
supersymmetric with respect to the Poincar\'e supersymmetric
transformations (\ref{mpea}) since it depends only on the spin
connection and $\delta\omega^{ab}=0$. This means that although the
Wigner-In\"on\"u contraction gives rise to a consistent theory, the
resulting Lagrangian, being decoupled of the vielbein and the
gravitino, is of little interest. The next order $l^{-1}$ in the
expansion (\ref{pokk}) is more interesting since it contains the
Poincar\'e invariant gravity Lagrangian. However, it is clear with
the use of the Fierz rearrangement (\ref{Fr}) that the expression at
the order $l^{-1}$ is not supersymmetric because of the absence of
the bosonic one-form field $b_{(2)}^{ab}$. The natural way to
introduce this field is to exploit the fact that one can always add
to the spin connection a tensor under the Lorentz group as
\begin{eqnarray}
\omega^{ab}\to \omega^{ab}-\frac{1}{l}b_{(2)}^{ab}. \label{splittt}
\end{eqnarray}
Apart from introducing the required bosonic field $b_{(2)}^{ab}$,
the splitting (\ref{splittt}) has two other important consequences.
Firstly, it prescribes the correct supersymmetric transformation of
the bosonic field $b_{(2)}^{ab}$,
\begin{eqnarray}
{\delta}\omega^{ab}-\frac{1}{l}\delta
b_{(2)}^{ab}=-\frac{1}{l}\left(\bar{\epsilon}\Gamma^{ab}\psi\right)\Longrightarrow
{\delta}\omega^{ab}=0,\qquad \delta
b_{(2)}^{ab}=\left(\bar{\epsilon}\Gamma^{ab}\psi\right). \label{bab}
\end{eqnarray}
Secondly, the Lorentz Chern-Simons form ${\mathcal L}^{\star}$
brings now a contribution at the order $l^{-1}$ in the expansion
given by
$$
{\mathcal L}^{\star}(\omega-\frac{1}{l}b_{(2)})={\mathcal
L}^{\star}(\omega)-\frac{1}{l}\mbox{Tr}\left(
R\!\!\!\!\slash^{4k+1}b_{(2)}\!\!\!\!\!\!\!\!\!\slash\;\;\;\;\right)+o(l^{-2}).
$$
Combining this expression together with (\ref{pokk}) shows that the
expression at the order $l^{-1}$ in the expansion becomes precisely
the supersymmetric action associated to the maximal super-Poincar\'e
algebra (\ref{pok}).

Hence, the connection between the AdS and Poincar\'e supersymmetric
theories has been realized through a standard Wigner-In\"on\"u
contraction supplemented by a splitting of the spin connection.
These are the two basic ingredients of the expansion method at the
algebraic level which is described as follows. Firstly, one
trivially extend the super AdS algebra with Lorentz generators
$T_{ab}$ satisfying $\left[T_{ab},
T_{cd}\right]=-T_{ac}\eta_{bd}+\cdots$ and, secondly one perform the
following contraction
\begin{eqnarray}
J_{ab}\to {J}_{ab}-T_{ab},\quad Z_{ab}= \frac{1}{l} T_{ab},\quad
P_{a}\to lP_a,\qquad
 {Z}_{a_1\cdots a_{p^{\prime}}}\to l Z_{a_1\cdots
a_{p^{\prime}}},\qquad {Q}\to \sqrt{l}\, Q,\label{cont}
\end{eqnarray}
where $l$ is the parameter of the expansion. In the limit $l\to
\infty$, the minimal extension of the AdS algebra (\ref{superAdS})
expressed in terms of the generators $J_{ab}, P_{a}, Z_{ab},
Z_{a_1\cdots a_{p^{\prime}}}$ and $Q_{\alpha}$ becomes precisely the
maximal supersymmetric extension of the Poincar\'e algebra
(\ref{maj}). The first operation is compatible with the splitting of
the spin connection while the second operation is nothing but a
standard Wigner-In\"on\"u contraction.

The lack of supersymmetry of the Lagrangian ${\mathcal L}^{(1)}$ in
the expression (\ref{pokk}) is due to the presence of the
Lorentz-Chern-Simons form. This form defined only in odd dimensions
$d=4k-1$ (which includes the odd dimensions allowing Majorana
spinors) is part of the Pontryagin-Chern-Simons form which is
required in order to further extend the AdS symmetry into
supersymmetry without duplicating the field content of the theory
\cite{HOR}. Indeed, the action of the standard Wigner-In\"on\"u
contraction on the dynamical fields (\ref{resmathcale}) also affect
the original supersymmetric transformations (\ref{mpae}). Indeed,
the gauge parameter $\epsilon$ must be rescaled as $\epsilon \to
1/\sqrt{l}\,\epsilon$ and, as a consequence the supersymmetric
transformations (\ref{mpae}) are split into two different orders. In
particular the variation of the spin connection gets a contribution
of the first order,
\begin{eqnarray}
\delta\omega^{ab}=-\frac{1}{l}\left(\bar{\epsilon}\Gamma^{ab}\psi\right)\Longrightarrow
\delta^{(0)} \omega^{ab}=0, \qquad  \delta^{(1)}
\omega^{ab}=-\frac{1}{l}\left(\bar{\epsilon}\Gamma^{ab}\psi\right)
\label{omgemepa}
\end{eqnarray}
while for the remaining dynamical fields we have
\begin{eqnarray}
&&\delta^{(0)} e^a=\left(\bar{\epsilon}\Gamma^a\psi\right),\quad
\delta^{(1)} e^a=0\nonumber\\
&&\delta^{(0)}b_{(p^{\prime})}^{a_1\cdots
a_{p^{\prime}}}=(\bar{\epsilon}\Gamma^{a_1\cdots
a_{p^{\prime}}}\psi),\quad \delta^{(1)}b_{(p^{\prime})}^{a_1\cdots
a_{p^{\prime}}}=0\nonumber\\
&&\delta^{(0)}\psi=D(\omega)\epsilon,\quad
\delta^{(1)}\psi=(e_a\Gamma^a+\sum_{p^{\prime}}b_{(p^{\prime})}^{a_1\cdots
a_{p^{\prime}}}\Gamma_{a_1\cdots a_{p^{\prime}}})\epsilon
\label{susyrescale}
\end{eqnarray}
Hence the variation of the original supersymmetric Lagrangian
(\ref{pokk}) under the supersymmetric transformations (\ref{mpae})
gives a series in powers of $l^{-1}$ in which each order is a total
derivative, i.e.
\begin{eqnarray}
\delta {\mathcal L}_{AdS}^{\mbox{\tiny{susy}}}= &&[\delta^{(0)}
{\cal L}^{(0)}]+\frac{1}{l} [\delta^{(0)} {\cal
L}^{(1)}+\delta^{(1)} {\cal
L}^{(0)}] +o(l^{-2})\nonumber\\
=&&d\Sigma_0+\frac{1}{l}d\Sigma_1+o(l^{-2}). \label{mas}
\end{eqnarray}
From this expression, it is clear that the Lagrangian ${\cal
L}^{(1)}$ is not supersymmetric w. r. t. the transformations
$\delta^{(0)}$ since the variation of the Lorentz-Chern-Simons form
$\delta^{(1)} {\cal L}^{(0)}$ does not vanish and is not a surface
term. This is because ${\cal L}^{(0)}$ depends on the spin
connection and the variation of the latter has a contribution at the
first order (\ref{omgemepa}). To circumvent this problem, it would
be sufficient if the variation $\delta^{(1)}\omega^{ab}$ vanished
identically, and this is precisely what we have done previously. It
is interesting to note that in doing so, the transformations
$\delta^{(0)}$ given in (\ref{susyrescale}) together with the
transformation of the extra field arising from the splitting of the
spin connection (\ref{bab}) are precisely the gauge transformations
associated to the maximal extension of the ${\mathcal N}=1$
super-Poincar\'e algebra (\ref{mpea}). Hence the expansion method in
order to make supersymmetric the next order naturally brings in the
gauge structure of the maximal extension of the ${\mathcal N}=1$
super-Poincar\'e algebra.

In this analysis, we have been concerned by the $(4k+2)$th Chern
character whose potential Chern-Simons is given by the Lagrangian
${\mathcal L}_{\tiny{\mbox{AdS}}}^{\tiny{\mbox{susy}}}$,
(\ref{AdSsusyaction}). However, in dimension $d=3+4k$, there exist
many other Chern character forms such that
$\mbox{STr}(F^{4k})\,\mbox{STr}(F^{2})$ or
$\left[\mbox{STr}(F^{2})\right]^{1+2k}$. A natural question to ask
is whether our conclusions depend on the choice of the Chern
character. However, it is tedious but straightforward to see that
any linear combination of all the Chern characters, namely
$$
\alpha_1
\mbox{STr}(F^{4k+2})+\alpha_2\mbox{STr}(F^{4k})\,\mbox{STr}(F^{2})+\cdots
+\alpha_p \left[\mbox{STr}(F^{2})\right]^{1+2k},
$$
will lead through the same procedure to the same conclusions: the
Wigner-In\"on\"u contraction will give a Lagrangian decoupled from
the vielbein and in order to make the next order supersymmetric, the
maximal extension of the ${\mathcal N}=1$ super-Poincar\'e algebra
will naturally emerge. The resulting Lagrangian will be the one
derived in the previous section (\ref{pok}) up to some additional
terms decoupled from the vielbein that are supersymmetric by
themselves.

\section{Discussion}
In odd dimensions allowing Majorana spinors $d=3+8k$, we have
considered a gravitational Lagrangian given by the dimensional
continuation of the Euler density and possessing the local
Poincar\'e invariance. We have constructed the local supersymmetric
extension of this Poincar\'e invariant gravity Lagrangian and we
have shown that the resulting action can be viewed as a gauge theory
for the maximal extension of the ${\mathcal N}=1$ super-Poincar\'e
algebra. We have addressed the same question for the AdS invariant
gravity for which we have constructed its local supersymmetric
extension for the minimal extension of the AdS algebra. In these
particular dimensions, the maximal super-Poincar\'e algebra has more
generators than the minimal super AdS algebra, and hence it is clear
that these two super algebras can not be linked through a process
like the Wigner-In\"on\"u contraction that does not increase or
decrease the number of generators. This is the reason for which we
have taken advantage of the expansion method developed in
\cite{deAzcarraga:2002xi} which permits to link Lie algebras of
different dimensions. More precisely, the maximal extension of the
${\mathcal N}=1$ super-Poincar\'e algebra can be obtained through
the expansion from the minimal super AdS algebra. This
correspondence at the algebraic level has been extended at the level
of the corresponding supersymmetric actions. In fact, we have
obtained the local supersymmetric extension of this Poincar\'e
invariant gravity Lagrangian from the minimal super AdS gravity
Lagrangian by a process compatible with the expansion method. In
doing so, we have pointed out that the necessity of expanding the
minimal super AdS algebra rather than considering a standard
Wigner-In\"on\"u is a direct consequence of the presence of the
Lorentz-Chern-Simons form (\ref{LCS}) . This form defined only in
odd dimensions $d=4k-1$ is part of the Pontryagin-Chern-Simons form
which is required in order to further extend the AdS symmetry into
supersymmetry without duplicating the field content of the theory
\cite{HOR}. It would be interesting to see what is the correct
correlation between the presence of the Pontryagin-Chern-Simons form
and the necessity of expanding the algebra rather than operating a
standard contraction. We have also shown that our conclusions do not
depend on the choice of the Chern character and, any linear
combination of the different Chern characters will lead to the same
conclusion.

In $d=3+8k$, the expansion has been realized at the first order.
This has permitted to supersymmetrize the Poincar\'e invariant
gravity Lagrangian. At the order $1+8k$, it will appear the standard
Einstein-Hilbert Lagrangian and hence, it will be interesting to see
whether one can construct a consistent supersymmetric theory at this
order through the same procedure. Note that starting from the AdS
invariant gravity Lagrangian in the absence of supersymmetry, it is
possible to deform the theory through the expansion of the Lie
algebra and get a system consisting of the Einstein-Hilbert action
plus nonminimally coupled matter \cite{Edelstein:2006se}.

In the expansion procedure, we have pointed out that the resulting
supersymmetric extension of the Poincar\'e invariant gravity is a
gauge theory for the maximal extension of the super-Poincar\'e
algebra. However, there exist supersymmetric extensions of the
Poincar\'e invariant gravity that are gauge systems for some
subalgebras of the maximal extension of the super-Poincar\'e
algebra. It is clear from our analysis that these theories can not
be reached through the expansion. It will be interesting to explore
if there exists other process that will realize this task. In some
case, the subalgebra of the maximal extension of the
super-Poincar\'e algebra has the same dimension than the minimal
super AdS algebra. However, although these algebras can be put in
correspondence through the Wigner-In\"on\"u contraction, this is not
true for the corresponding supersymmetric actions. Hence, it would
be interesting to determine whether generalizations of the
Wigner-In\"on\"u contraction described in \cite{EWW} or in
\cite{IRP} can be useful for generating other supersymmetric
theories than those obtained through the expansion method.

Finally, we have only been concerned with the odd dimensions
allowing Majorana spinors. A natural extension of this work will
consist in considering all the odd dimensions and to realize an
exhaustive analysis \cite{HR}.

\bigskip

\acknowledgments  We thank E. Ay\'on-Beato, J. Edelstein, R.
Troncoso and J. Zanelli for useful discussions and P. Desrosiers for
a careful reading of this manuscript. This work is partially
supported by grants 1051084 and 1060831 from FONDECYT.



\begin{thebibliography}{99}
\bibitem{LAN} C. Lanczos, Ann. Math {\textbf 39}, 842 (1938).

\bibitem{LOV} D. Lovelock, J. Math. Phys. {\textbf 12}, 498 (1971).

\bibitem{TEZ} C. Teitelboim and J. Zanelli, Class. Quant. Grav. {\textbf 4}, L125 (1987).

\bibitem{ZW} B. Zwiebach, Phys. Lett {\textbf B156}, 315 (1985).

\bibitem{MSSK} N.~Marcus and J.~H.~Schwarz, Nucl.\ Phys.\ B {\bf 228}, 145
(1983) ; S.~Deser and J.~H.~Kay, Phys.\ Lett.\ B {\bf 120}, 97
(1983); S. Deser, in {\it Quantum Theory of Gravity: Essays in Honor
of the 60th Birthday of Bryce S. deWitt}, edited by S. M.
Christensen (Adam Hilger, London, 1984).

\bibitem{BTZ} M. Ba\~nados, R. Troncoso and J. Zanelli, Phys.Rev.D {\textbf 54}, 2605 (1996)

\bibitem{JZ}J.~Zanelli, {\it Lecture notes on Chern-Simons
(super-)gravities}, arXiv:hep-th/0502193; Braz.\ J.\ Phys.\  {\bf
30}, 251 (2000); J.~D.~Edelstein and J.~Zanelli, J.\ Phys.\ Conf.\
Ser.\  {\bf 33}, 83 (2006).


\bibitem{HOT} M.~Hassa\"{\i}ne, R.~Olea and R.~Troncoso, Phys.\ Lett. {\bf B599}, 111 (2004)


\bibitem{HTZ} M.~Hassa\"{\i}ne, R.~Troncoso and J.~Zanelli, Phys. Lett. \textbf{B596}, 132
(2004); Proc.\ Sci.\  {\bf WC2004}, 006 (2005).


\bibitem{CJS} E. Cremmer, B. Julia and J. Scherk,
Phys. Lett. \textbf{B76}, 409 (1978).



\bibitem{vanProeyen} A.~Van Proeyen, {\it Tools for supersymmetry}, arXiv:hep-th/9910030.


\bibitem{deAzcarraga:2002xi} J.~A.~de Azcarraga, J.~M.~Izquierdo, M.~Picon and O.~Varela,
Nucl.\ Phys.\ B {\bf 662}, 185 (2003);  Int.\ J.\ Theor.\ Phys.\
{\bf 46}, 2738 (2007).

\bibitem{HS} M.~Hatsuda and M.~Sakaguchi, Prog.\ Theor.\ Phys.\  {\bf 109}, 853
(2003).






\bibitem{S} I. E. Segal, Duke Math. J. {\textbf 18}, 221 (1951).


\bibitem{HTZ2} M.~Hassa\"{\i}ne, R.~Troncoso and J.~Zanelli,
{\it Emergence of the M-algebra from eleven-dimensional AdS
supergravity}, preprint CECS-PHY-06/08.





\bibitem{AT}A. Ach\'{u}carro and P.K. Townsend, Phys. Lett. \textbf{B180}, 89 (1986).


\bibitem{GTW}A.~Giacomini, R.~Troncoso and S.~Willison, Class.\ Quant.\ Grav.\  {\bf 24}, 2845
(2007).

\bibitem{CHA} A. H. Chamseddine, Nucl. Phys. \textbf{B346}, 213 (1990).

\bibitem{TZ} R. Troncoso and J. Zanelli, Phys.Rev.D {\textbf 58}, 101703 (1998);
Int.J.Theor.Phys. {\textbf 38}, 1181 (1999);  Class.Quant.Grav.
{\textbf 17}, 4451 (2000).






\bibitem{vanHolten:1982mx} J.~W.~van Holten and A.~Van Proeyen, J.\ Phys.\ A {\bf 15}, 3763 (1982).









\bibitem{HOR} P.~Horava, Phys. Rev. D{\textbf 59}, 046004 (1999).



\bibitem{Edelstein:2006se}
J.~D.~Edelstein, M.~Hassaine, R.~Troncoso and J.~Zanelli, Phys.\
Lett.\  B {\bf 640}, 278, (2006).


\bibitem{EWW} E. Weimar-Woods, J. Math. Phys. {\bf 36}, 4519 (1995); Rev. Math. Phys., {\bf 12}, 1505,
(2000).

\bibitem{IRP} F.~Izaurieta, E.~Rodriguez and P.~Salgado, J.\ Math.\ Phys.\  {\bf 47}, 123512 (2006).


\bibitem{HR} M. Hassaine and M. Romo, work in progress.




\end{thebibliography}
\end{document}